\documentclass[12pt,reqno]{amsart}
\usepackage{graphicx}
\usepackage{amssymb,amsmath}
\usepackage{amsthm}
\usepackage{color}
\usepackage[pdf]{pstricks}
\usepackage{hyperref}
\usepackage{empheq}
\usepackage{pgfplots}

\usepackage{amssymb}
\usepackage{epsfig}
\usepackage{extarrows}
\usepackage{amsfonts}
\usepackage{graphicx}
\usepackage{caption}
\usepackage{subcaption}
\usepackage{tikz}
\usepackage{color}
\usepackage{subcaption}
\usepackage{float}

\begin{document}
	
\title{Algebraic solitons in the massive Thirring model}

\author{Jiaqi Han}
\address[J. Han]{Department of Mathematics and Statistics, McMaster University, Hamilton, Ontario, Canada, L8S 4K1}
\email{han90@mcmaster.ca}

\author{Cheng He}
\address[C. He]{School of Mathematics and Statistics, Ningbo University, Ningbo 315211, People's Republic of China}
\email{1811071003@nbu.edu.cn}

\author{Dmitry E. Pelinovsky}
\address[D. E. Pelinovsky]{Department of Mathematics and Statistics, McMaster University, Hamilton, Ontario, Canada, L8S 4K1}
\email{dmpeli@math.mcmaster.ca}

\begin{abstract}
	We present exact solutions describing dynamics of two algebraic solitons in the massive Thirring model. Each algebraic soliton corresponds to a simple embedded eigenvalue in the Kaup--Newell spectral problem and attains the maximal mass among the family of solitary waves traveling with the same speed. By coalescence of speeds of the two algebraic solitons, we find a new solution for an algebraic double-soliton which corresponds to a double embedded eigenvalue. We show that the double-soliton attains the double mass of a single soliton and describes a slow interaction of two identical algebraic solitons.
\end{abstract}   

\maketitle


\section{Introduction}

Algebraic solitons are traveling solitary waves with the power rather than exponential decay rate at infinity. Such solutions are common for integrable nonlinear equations with nonlocal terms such as the Benjamin--Ono and Kadomtsev--Petviashvili equations, where they are associated with isolated eigenvalues of the linear Lax equations \cite{AC90}. However, algebraic solitons are special for local integrable nonlinear equations since they arise as the limiting points in the family of exponential solitons and they are associated 
with embedded eigenvalues in the continuous spectrum of the linear Lax equations \cite{KPR06,PG97}. Physically relevant examples of the algebraic solitons 
as special limits of exponential solitons appear in the modified Korteweg--de Vries equation \cite{Akh16,HeAlg1}, the derivative nonlinear Schr\"{o}dinger equation \cite{Ling,HeAlg2,HeAlg3}, and the nonlinear Dirac equation \cite{Guo23}. 

This work is devoted to the algebraic solitons in the massive Thirring model (MTM) written in laboratory coordinates as
\begin{equation}
\label{MTM}
\left\{\begin{array}{l}
\mathrm{i} (u_t + u_x) + v = |v|^2 u \\
\mathrm{i} (v_t - v_x) + u=|u|^2 v
\end{array}\right.
\end{equation} 
where  $(u,v) \in \mathbb{C}^2$ and subscripts denote partial derivatives in 
$(x,t) \in \mathbb{R}^2$. The MTM system (\ref{MTM}) is a prototypical Dirac equation which belongs to the class of integrable equations associated with the Kaup--Newell (KN) spectral problem \cite{KN76,KM77}. 

Stability of algebraic solitons is a notoriously difficult mathematical problem, where every method of nonlinear analysis known in the theory of integrable systems fails. Coercivity of the energy function required for the proof of Lyapunov stability holds for exponential solitons \cite{PS14} but fails for algebraic solitons because the spectral gap between the zero eigenvalue and the continuous spectrum in the linearized MTM system closes up in the limit to the algebraic soliton. Stability of exponential solitons in the MTM system can be proven with the Darboux transformation \cite{CPS16} which allows us to construct exponential solitons from an isolated eigenvalue of the KN spectral problem. However, the Darboux transformation does not allow us to obtain algebraic solitons because the embedded eigenvalue has to be defined inside the continuous spectrum of the KN spectral problem, where both eigenfunctions are bounded. Finally, the inverse scattering transform (IST) method requires fast spatial decay of solutions of the MTM system at infinity in order to ensure smoothness properties of the scattering data and solvability of the associated Riemann--Hilbert problems \cite{Cheng23,PS19}. Algebraic solitons decay too slowly and violate the requirements of the fast spatial decay.

Due to these limitations, the main purpose of this paper is to explore direct methods of solutions of the MTM system (\ref{MTM}) and to study interaction of two algebraic solitons.  Hirota's bilinear formulation of the MTM system (\ref{MTM}) was recently developed in \cite{Feng1} to obtain exponential multi-solitons. By using the analytical expressions for two exponential solitons, we obtain the exact solutions for two algebraic solitons which scatter fast from each other with two different wave speeds. In the limit when the wave speeds coincide, we obtain the double-soliton solution which describes a slow interaction of two identical algebraic solitons. 

There has been a recent spike in the study of rational solutions of the integrable systems in the context of rogue wave dynamics \cite{Yang1,Yang2,Yang3}. Similar rational solutions of the MTM system for rogue waves were studied in \cite{Feng2,He,Ye}, where they appear on the constant, modulationally unstable background. Compared to these solutions, our rational solution for the algebraic double-soliton is not a rogue wave since it describes two algebraic solitons at the trivial, modulationally stable background. To the best of our knowledge, the double-soliton solution is derived in the MTM system (\ref{MTM}) for the first time. Its existence suggests that a hierarchy of higher-order rational solutions exists in the MTM system (\ref{MTM}) which has not been explored yet. Similar solutions for the algebraic double solitons in the derivative NLS equation were constructed in \cite{Ling} by using generalized Darboux transformation and in \cite{HeAlg3} by taking the limit of the exponential double-solitons obtained from the double-fold Darboux transformation. 

Double solitons traveling with nearly the same speed correspond to 
double eigenvalues of the KN spectral problem. The IST method was recently employed to construct exponential double-solitons of the derivative NLS equation and related equations \cite{Yan,He2,He1,Zhang}. Such solutions are related to double isolated 
eigenvalues or, equivalently, to double poles in solutions of the Riemann--Hilbert problems. The method does not work for the double embedded 
eigenvalues. In another work \cite{LiPel}, we will obtain the exponential 
double-solitons of the MTM system (\ref{MTM}) by using the IST method and show that these solutions degenerate into the algebraic double-solitons in the limit when the double isolated eigenvalue becomes embedded in the continuous spectrum of the KN spectral problem. This provides the first nontrivial example of embedded eigenvalues of higher algebraic multiplicity for the KN spectral problem, the possibility of which was first predicted 20 years ago in \cite{KPR06}. 

Our results suggest stability of the traveling algebraic solitons in the time evolution of the MTM system (\ref{MTM}). This conclusion agrees with the perturbation theory for embedded eigenvalues of the KN spectral problem 
developed in \cite[Sections 6-7]{KPR06}, which suggests that the simple embedded eigenvalues for algebraic solitons are generally ejected from the continuous spectrum to become simple isolated eigenvalues for the exponential solitons with nearly selected speed and frequency. A rigorous proof of orbital stability 
of the traveling algebraic soliton is still an open problem for the MTM system (\ref{MTM}). 

The rest of this paper is organized as follows. The main results are presented in Section \ref{sec2}. The computational proofs are elaborated in Section \ref{sec3} where we obtain a new parameterization of the exponential two-soliton solutions of the MTM system (\ref{MTM}) and then take the limits to the algebraic two-soliton solutions with different speeds and to the algebraic double-soliton of the same speed. Section \ref{sec4} emphasizes further directions which may be undertaken from the outcomes of our work.

\section{Main results}
\label{sec2}

To simplify presentation of soliton solutions of the MTM system (\ref{MTM}), we shall use the basic symmetries of this Hamiltonian system. These include the translational and rotational symmetries 
\begin{equation}
\label{MTM-symm}
\left[ \begin{matrix}
u(x,t) \\
v(x,t)  
\end{matrix} \right]  \quad \mapsto \quad 
\left[ \begin{matrix}
u(x+x_0,t+t_0) e^{\mathrm{i} \theta_0} \\
v(x+x_0,t+t_0) e^{\mathrm{i} \theta_0}
\end{matrix}\right],
\qquad x_0, t_0, \theta_0 \in \mathbb{R},
\end{equation}
as well as the Lorentz symmetry
\begin{equation}
\label{MTM-Lorentz}
\left[ \begin{matrix}
u(x,t) \\
v(x,t)  
\end{matrix} \right]  \;\; \mapsto \;\; 
\left[ \begin{matrix}
\left( \frac{1 - c}{1 + c} \right)^{1/4} u\left( \frac{x + ct}{\sqrt{1-c^2}}, \frac{t + cx}{\sqrt{1-c^2}} \right) \\
\left( \frac{1 + c}{1 - c} \right)^{1/4} v\left( \frac{x + ct}{\sqrt{1-c^2}}, \frac{t + cx}{\sqrt{1-c^2}} \right)
\end{matrix}\right], 
\qquad c \in (-1,1).
\end{equation}
Without loss of generality, each solution of the MTM system (\ref{MTM}) can be extended with three translational parameter in (\ref{MTM-symm}) and the speed parameter $c \in (-1,1)$ in (\ref{MTM-Lorentz}).

A normalized family of exponential solitons of the MTM system (\ref{MTM}) is given by the standing wave solutions of the form 
\begin{equation}
\label{MTM-solitary}
\left[ \begin{matrix}
u_{\rm sol}(x,t) \\
v_{\rm sol}(x,t)  
\end{matrix} \right]  = 
\sin \gamma \left[ \begin{matrix} {\rm sech}\left(x \; \sin \gamma + \frac{\mathrm{i} \gamma}{2} \right) \\  {\rm sech}\left(x \; \sin \gamma - \frac{\mathrm{i} \gamma}{2} \right) \end{matrix} \right]  e^{\mathrm{i} t \cos \gamma}, \quad \gamma \in (0,\pi).
\end{equation}
A general family with two translational parameters and the speed parameter $c \in (-1,1)$ is obtained from the translational and Lorentz symmetry given by (\ref{MTM-symm}) and (\ref{MTM-Lorentz}). 

The only parameter $\gamma \in (0,\pi)$ in (\ref{MTM-solitary}) defines 
the frequency parameter $\omega := \cos(\gamma)$ of the exponential solitons. 
The frequency $\omega$ is chosen in the gap $(-1,1)$ of the frequency spectrum  of the linear Dirac operator 
$$
\mathcal{D} := \left[ \begin{matrix} \mathrm{i} \partial_x & 1 \\ 1 & -\mathrm{i} \partial_x \end{matrix} \right],
$$
which determines the time evolution of the MTM system (\ref{MTM}). This is one of the reasons why Dirac solitons are sometimes called the gap solitons \cite{Pel11}. 

The limits $\omega \to \pm 1$  are referred to as the nonrelativistic limits of the MTM system (\ref{MTM}). It is well-known (see, e.g., \cite{Borreli,BC19,Guo21}) that the nonlinear Dirac equations such as the MTM system (\ref{MTM}) can be reduced to the focusing NLS equation as $\omega \to 1$ and to the defocusing NLS equation as $\omega \to -1$. The normalized form for the two NLS equations is given by 
\begin{equation}
\label{NLS}
\mathrm{i} \psi_t + \psi_{xx} + \sigma |\psi|^2 \psi = 0, \qquad \sigma = {\rm sgn}(\omega) = \pm 1.
\end{equation}
The family (\ref{MTM-solitary}) reduces to the small-amplitude, long-scale, ${\rm sech}$-shaped soliton of the focusing NLS equation (\ref{NLS}) with $\sigma = +1$ as $\omega \to 1$ ($\gamma \to 0$) and to the finite-amplitude, finite-scale, algebraic soliton 
\begin{equation}
\label{MTM-algebraic}
\gamma = \pi : \quad 
\left[ \begin{matrix}
u_{\rm alg}(x,t) \\
v_{\rm alg}(x,t)  
\end{matrix} \right]  = 
\left[ \begin{matrix} \displaystyle  \frac{2}{1+2 \mathrm{i} x} \\ \displaystyle  \frac{2}{1-2 \mathrm{i} x} \end{matrix} \right]  e^{-\mathrm{i} t}
\end{equation}
as $\omega \to -1$ ($\gamma \to \pi$). Note that the NLS equations (\ref{NLS}) holds in the limit of small amplitudes and hence the algebraic soliton (\ref{MTM-algebraic}) does not satisfy the reduction to the defocusing NLS equation (\ref{NLS}) with $\sigma = -1$ as $\omega \to -1$.

The algebraic soliton (\ref{MTM-algebraic}) has the largest mass among 
the exponential solitons in the family (\ref{MTM-solitary}), where the mass 
for the MTM system (\ref{MTM}) is defined by 
\begin{equation}
Q(u,v) := \int_{\mathbb{R}} (|u|^2 + |v|^2) dx.
\end{equation}
It follows from (\ref{MTM-solitary}) that 
\begin{equation*}
|u(x,t)|^2 + |v(x,t)|^2 = \frac{4 \sin^2 \gamma}{\cos \gamma + \cosh(2 x \; \sin \gamma)}, 
\end{equation*}
which implies that $Q(u_{\rm sol},v_{\rm sol}) = 4 \gamma$ with the largest mass at $Q(u_{\rm alg},v_{\rm alg}) = 4 \pi$. 

Let us now present the main results of this paper. First, we have obtained the exact formula for the algebraic two-soliton solution of the MTM system (\ref{MTM}). The exact formula can be written in the form:
\begin{equation}
\label{u-two-sol}
u(x,t) = -\frac{2 \mathrm{i} \delta_1^{-1/2} e^{-\mathrm{i} T_1} \left[  2 X_2 + \mathrm{i} - \frac{4 \mathrm{i} \delta_1}{\delta_1 - \delta_2} \right] + 2 \mathrm{i} \delta_2^{-1/2} e^{-\mathrm{i} T_2} \left[  2 X_1 + \mathrm{i} + \frac{4 \mathrm{i} \delta_2}{\delta_1 - \delta_2} \right]}{(2 X_1 - \mathrm{i}) (2 X_2 - \mathrm{i}) + \frac{4 \sqrt{\delta_1 \delta_2}}{(\delta_1 - \delta_2)^2} 
	\left[ \sqrt{\delta_1} e^{\frac{\mathrm{i}}{2} (T_1 - T_2)} - \sqrt{\delta_2} e^{-\frac{\mathrm{i}}{2} (T_1 - T_2)} \right]^2}
\end{equation}
and 
\begin{equation}
\label{v-two-sol}
v(x,t) = \frac{2 \mathrm{i} \delta_1^{1/2} e^{-\mathrm{i} T_1} \left[  2 X_2 - \mathrm{i} - \frac{4 \mathrm{i} \delta_2}{\delta_1 - \delta_2} \right] 
	+ 2 \mathrm{i} \delta_2^{1/2} e^{-\mathrm{i} T_2} \left[  2 X_1 - \mathrm{i} + \frac{4 \mathrm{i} \delta_1}{\delta_1 - \delta_2} \right]}{(2 X_1 + \mathrm{i}) (2 X_2 + \mathrm{i}) + \frac{4 \sqrt{\delta_1 \delta_2}}{(\delta_1 - \delta_2)^2} 
	\left[ \sqrt{\delta_1} e^{-\frac{\mathrm{i}}{2} (T_1 - T_2)} - \sqrt{\delta_2} e^{\frac{\mathrm{i}}{2} (T_1 - T_2)} \right]^2}
	\end{equation}
where 
\begin{align*}
X_j = \frac{x + c_j t}{\sqrt{1-c_j^2}} + x_j, \quad
T_j = \frac{t + c_j x}{\sqrt{1-c_j^2}} + t_j, \quad  c_j = \frac{\delta_j^2-1}{\delta_j^2+1}
\end{align*}
with the parameters $\delta_{1,2} > 0$ such that $\delta_1 \neq \delta_2$ and translational parameters $x_{1,2} \in \mathbb{R}$ and $t_{1,2} \in \mathbb{R}$. 

\begin{figure}[htb]
	\centering
	\includegraphics[width=7.5cm,height=6cm]{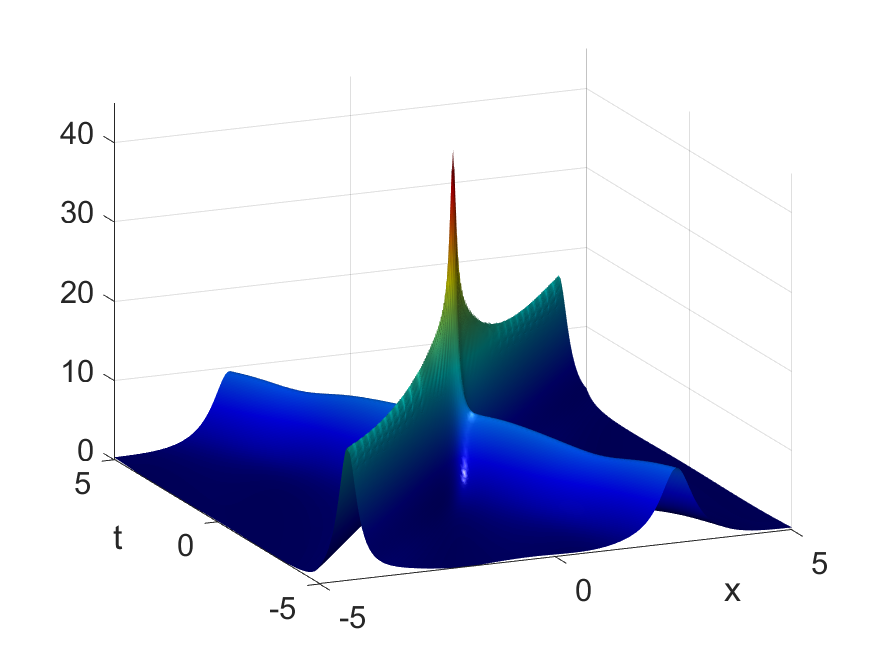} 
	\hspace{0.2cm}
	\includegraphics[width=7.5cm,height=6cm]{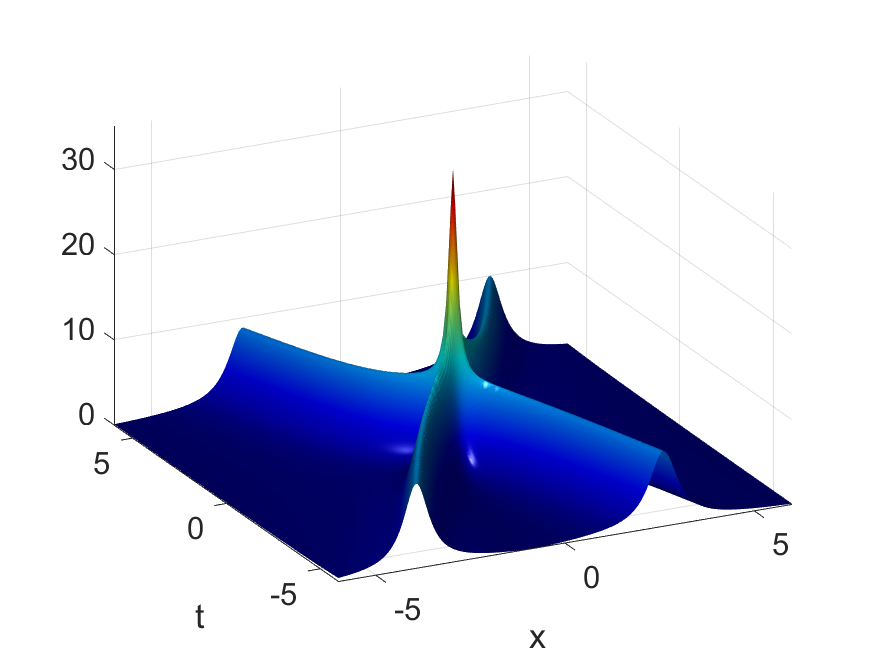} \\
	\includegraphics[width=7.5cm,height=6cm]{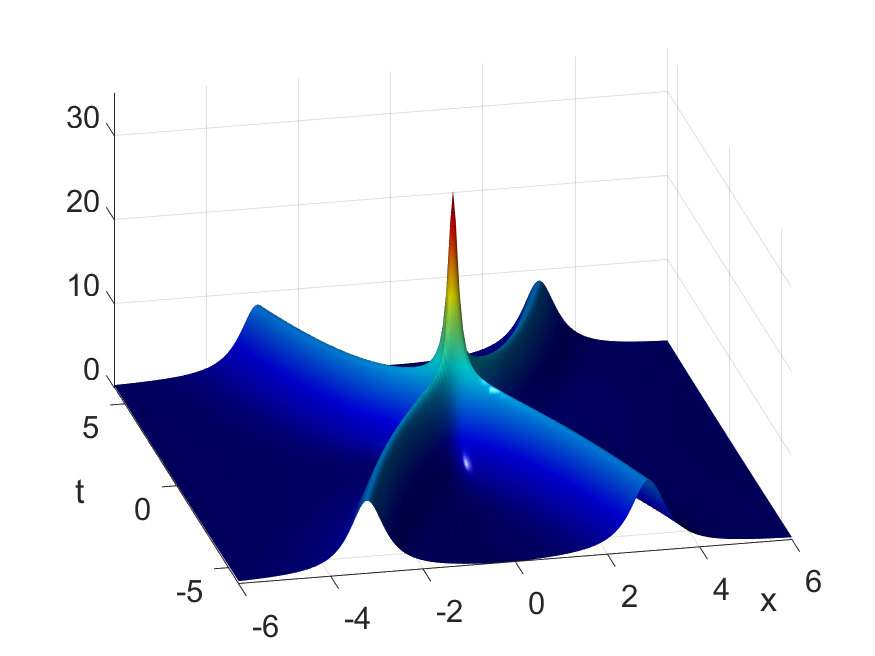} 
	\hspace{0.2cm}
	\includegraphics[width=7.5cm,height=6cm]{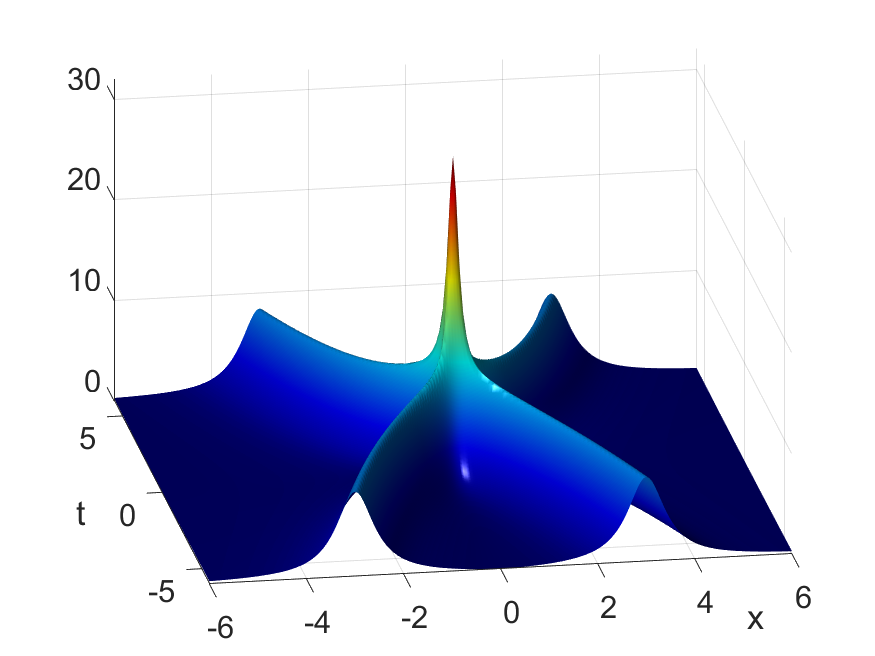}
	\caption{The solution surface for $|u|^2 + |v|^2$ versus $(x,t)$ for the family (\ref{u-two-sol}) and (\ref{v-two-sol}) with $x_1 = x_2 = t_1 = t_2 = 0$ and $\delta_1 = 1 + \varepsilon$, $\delta_2 = 1-\varepsilon$ with $\varepsilon = 0.75$ (top left), $\varepsilon = 0.5$ (top right), $\varepsilon = 0.25$ (bottom left), and $\varepsilon = 0.01$ (bottom right).}
	\label{fig:solitons}
\end{figure}

Figure \ref{fig:solitons} shows the solution surfaces which suggest that the algebraic two-soliton solution given by (\ref{u-two-sol}) and (\ref{v-two-sol}) describes scattering of two algebraic solitons. When the wave speeds $c_1$ and $c_2$ are very different from each other (top panels), the scattering is fast and the trajectories of the two solitons are almost straight lines. When the wave speeds approach to each other (bottom panels), the scattering becomes slow and the trajectories of the two solitons are curved near the soliton overlaping regions.  

In the limit $\delta_1, \delta_2 \to 1$ of the algebraic two-soliton solution given by  (\ref{u-two-sol}) and (\ref{v-two-sol}), we derived the following 
new rational solution to the MTM system (\ref{MTM}):
\begin{equation}
\label{MTM-double}
\left[ \begin{matrix}
u_{\rm double}(x,t) \\
v_{\rm double}(x,t)  
\end{matrix} \right]  = 
\left[ \begin{matrix} \displaystyle  \frac{4(-3 + 6 \mathrm{i} x - 12 x^2 - 8 \mathrm{i} x^3 - 12 t (2x - \mathrm{i}) - \mathrm{i}\beta)}{3 + 24 \mathrm{i} x - 24 x^2 + 32 \mathrm{i} x^3 - 16 x^4 + 48 t^2 + 2 \beta (2x - \mathrm{i}) } \\ \displaystyle  \frac{4(-3 - 6 \mathrm{i} x - 12 x^2 + 8 \mathrm{i} x^3 + 12 t (2x + \mathrm{i}) + \mathrm{i}\beta)}{3 - 24 \mathrm{i} x - 24 x^2 - 32 \mathrm{i} x^3 - 16 x^4 + 48 t^2 + 2 \beta (2x + \mathrm{i}) } \end{matrix} \right]  e^{-\mathrm{i} t}.
\end{equation}
where $\beta \in \mathbb{R}$ is a free parameter of the family. A general family with three translational parameters and the speed parameter $c \in (-1,1)$ is obtained from the translational and Lorentz symmetry given by (\ref{MTM-symm}) and (\ref{MTM-Lorentz}).

\begin{figure}[htb]
	\centering
	\includegraphics[width=7.5cm,height=6cm]{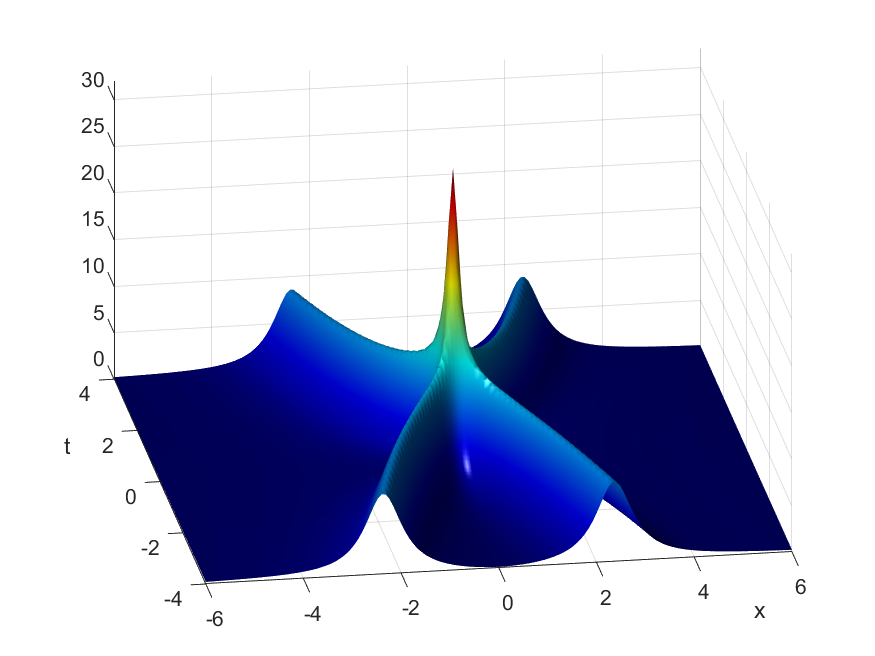} 
	\hspace{0.2cm}
	\includegraphics[width=7.5cm,height=6cm]{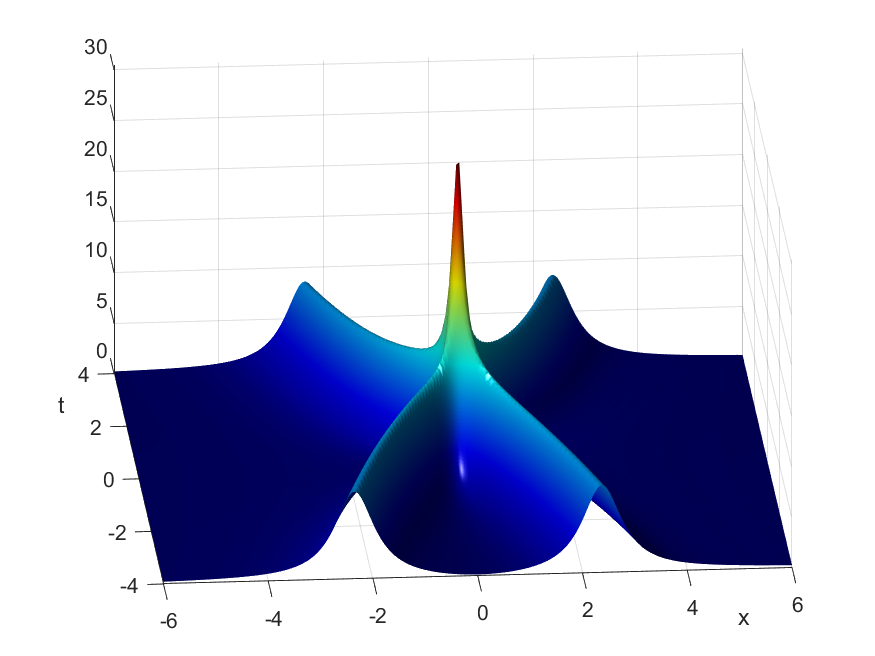} \\
	\includegraphics[width=7.5cm,height=6cm]{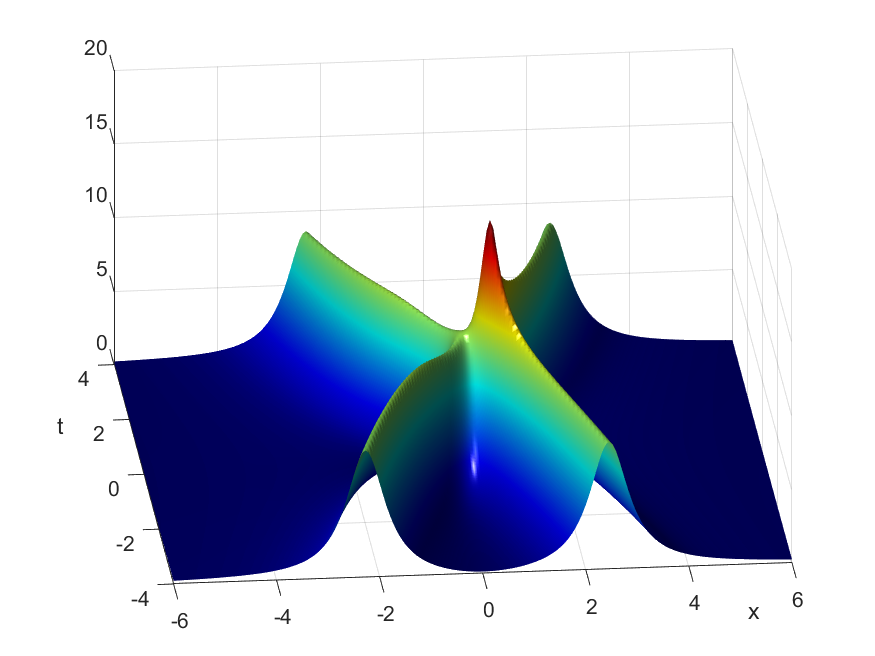} 
	\hspace{0.2cm}
	\includegraphics[width=7.5cm,height=6cm]{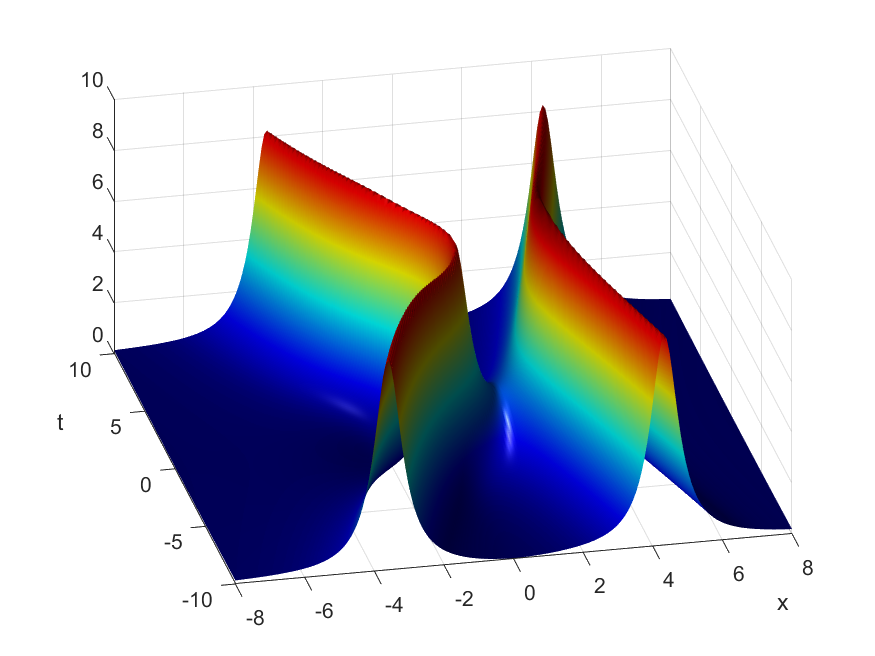}
	\caption{The solution surface for $|u|^2 + |v|^2$ versus $(x,t)$ for the family (\ref{MTM-degenerate}) with $\beta = 0$ (top left), $\beta = 1$ (top right), 
		$\beta = 10$ (bottom left), and $\beta = 100$ (bottom right).}
	\label{fig:uuvv}
\end{figure}

The algebraic double-soliton given by (\ref{MTM-double}) describes a slow scattering of two identical algebraic solitons. The parameter $\beta$ describes the distance between the two solitons. Figure \ref{fig:uuvv} illustrates the solution surface for $|u|^2 + |v|^2$ versus $(x,t)$ for the family of solutions (\ref{MTM-double}) with $\beta = 0, 1, 10, 100$. The solution with $\beta = 0$ is symmetric with the global maximum at $(0,0)$. Since $|u(0,0)|^2 + |v(0,0)|^2 = 32$ for (\ref{MTM-double}) 
and $|u(0,0)|^2 + |v(0,0)|^2 = 8$ for (\ref{MTM-algebraic}), 
the double-soliton has the quadruple magnification factor for the squared 
amplitudes compared to the single algebraic soliton. 

As $\beta$ increases, the symmetry is broken and the magnification factor becomes smaller. For sufficiently large $\beta$, the two solitons do not overlap but slowly scatter at a distance from each other. As $\beta \to \infty$, one soliton goes to infinity and the other soliton is located near the origin. Indeed, the family of solutions (\ref{MTM-double}) converges as $\beta \to \infty$ to a single algebraic soliton (\ref{MTM-algebraic}). 
We will prove that 
\begin{equation}
\label{double-mass}
Q(u_{\rm double},v_{\rm double}) = 8 \pi = 2 Q(u_{\rm alg},v_{\rm alg}),
\end{equation} 
which implies that the double-soliton (\ref{MTM-double}) has a double mass compared to the single algebraic soliton (\ref{MTM-algebraic}).

\begin{figure}[H]
	\centering
	\includegraphics[width=7.5cm,height=6cm]{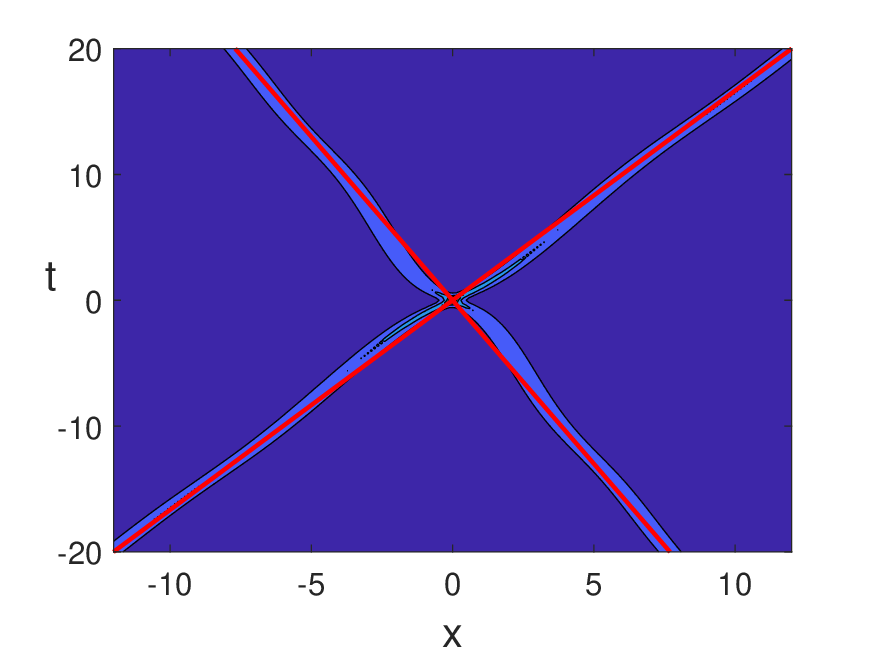} \vspace{0.2cm}
	\includegraphics[width=7.5cm,height=6cm]{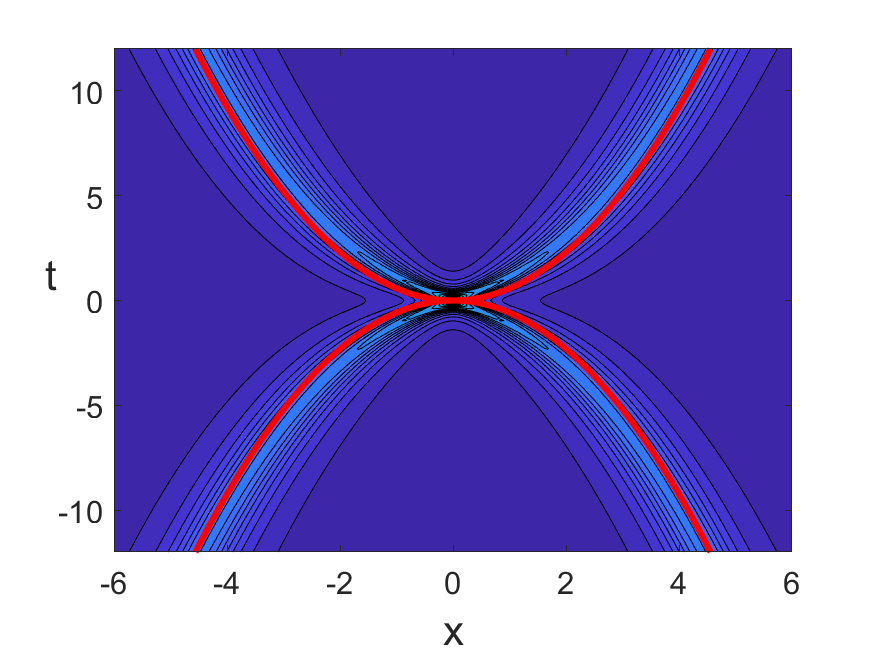} 
	\caption{The contour plots for the solution surfaces from Fig.  \ref{fig:solitons} with $\varepsilon = 0.5$ (left) and from Fig. \ref{fig:uuvv} with $\beta = 0$ (right). The red lines show the straight lines $x + c_1 t = 0$ and $x + c_2 t = 0$ (left) and the parabolas $x^2 = \pm 3t$ (right).}
	\label{fig-contour}
\end{figure}

For the fast scattering of two algebraic solitons given by (\ref{u-two-sol}) and (\ref{v-two-sol}), the algebraic solitons move along straight lines before and after interaction in the overlapping region. No phase shift arises as a result of the soliton interaction, which is a standard feature of algebraic multi-soliton solutions, see \cite{GPS93,S18}. This is illustrated on the contour plot of Figure \ref{fig-contour} (left panel), where we showed the solution from Figure \ref{fig:solitons} with $\varepsilon = 0.5$ together with the straight lines $x + c_1 t = 0$ and $x + c_2 t = 0$. On the other hand, the slow scattering of two identical solitons given by (\ref{MTM-double}) 
results in the solitons propagating along a curve on the $(x,t)$ plane. Figure \ref{fig-contour} (right panel) shows the solution from Figure \ref{fig:uuvv} with $\beta = 0$ together with the parabolas $x^2 = \pm 3t$. The free algebraic solitons would be standing waves with $c = 0$ but their slow interaction results in the dynamics along the trajectories at $x \approx \pm \sqrt{3t}$ as $t \to \infty$ with nonzero but asymptotically vanishing velocities $\pm \frac{\sqrt{3}}{2\sqrt{t}} \to 0$ as $t \to +\infty$.

The next section contains derivation of (\ref{u-two-sol})--(\ref{v-two-sol}) and (\ref{MTM-double}) as well as the proof of (\ref{double-mass}).

\section{Proof of the main results}
\label{sec3}

As a starting point, the MTM system (\ref{MTM}) is transformed to a system of bilinear equations by the following transformation \cite{Feng1}, 
\begin{equation}
\label{bilinear}
u=\frac{g}{\bar{f}}, \qquad v=\frac{h}{f},
\end{equation} 
where $\bar{f}$ is complex conjugate of $f$. Substituting (\ref{bilinear}) into (\ref{MTM}) yields the following system of bilinear equations for $f$, $h$, and $g$:
\begin{equation}
\label{Hirota}
\left. \begin{array}{r}
\mathrm{i} f (g_t + g_x) - \mathrm{i} g (f_t + f_x) + h \bar{f} = 0, \\
\mathrm{i} \bar{f} (h_t - h_x) - \mathrm{i} h (\bar{f}_t - \bar{f}_x) + g f = 0, \\
\mathrm{i} \bar{f} (f_x + f_t) - \mathrm{i} f (\bar{f}_t + \bar{f}_x) - |h|^2 = 0, \\
\mathrm{i} f (\bar{f}_t - \bar{f}_x) - \mathrm{i} \bar{f} (f_t - f_x) - |g|^2 = 0. 
\end{array}\right\}
\end{equation}
It was proven in \cite{Feng1} that the system (\ref{Hirota}) is satisfied by the following two-soliton solutions in the general form:
\begin{equation}
\left\{ \begin{array}{l}
f = 1 + c_{11} e^{\zeta_1 + \bar{\zeta}_1} + c_{12} e^{\zeta_1 + \bar{\zeta}_2} 
+ c_{21} e^{\bar{\zeta}_1 + \zeta_2} + c_{22} e^{\zeta_2 + \bar{\zeta}_2} 
+ c_{1212} e^{\zeta_1 + \bar{\zeta}_1 + \zeta_2 + \bar{\zeta}_2}, \\
h = \bar{\alpha}_1 e^{\zeta_1} + \bar{\alpha}_2 e^{\zeta_2} + 
c_{121} e^{\zeta_1 + \zeta_2 +\bar{\zeta}_1} + c_{122} 
e^{\zeta_1+\zeta_2+\bar{\zeta}_2}, \\
g = \frac{\mathrm{i} \bar{\alpha}_1}{p_1} e^{\zeta_1}+\frac{\mathrm{i} \bar{\alpha}_2}{p_2} e^{\zeta_2} - \frac{\mathrm{i} \bar{p}_1}{p_1 p_2} c_{121} e^{\zeta_1 + \zeta_2 + \bar{\zeta}_1} - \frac{\mathrm{i} \bar{p}_2}{p_1 p_2} c_{122} e^{\zeta_1+\zeta_2 +\bar{\zeta}_2},
\end{array} \right. 
\label{2-soliton}
\end{equation} 
where 
\begin{equation*}
\zeta_j = \frac{1}{2} \left(p_j+\frac{1}{p_j}\right) x + \frac{1}{2} \left(p_j-\frac{1}{p_j}\right) t
\end{equation*}
and
\begin{align*}
c_{i j} &=-\frac{\mathrm{i}  p_i \bar{\alpha}_i \alpha_j}{(p_i+\bar{p}_j)^2}, \\
c_{12 j} &= (p_1-p_2) \bar{p}_j \left[\frac{\bar{\alpha}_2 c_{1 j}}{p_1 (p_2 + \bar{p}_j)} - \frac{\bar{\alpha}_1 c_{2 j}}{p_2 (p_1+ \bar{p}_j)} \right], \\
c_{1212} &= |p_1-p_2|^2 \left[\frac{c_{11} c_{22}}{(p_1 + \bar{p}_2) (p_2 + \bar{p}_1)} - \frac{c_{12} c_{21}}{(p_1 + \bar{p}_1) (p_2 + \bar{p}_2)} \right],
\end{align*} 
whereas parameters $p_1, p_2, \alpha_1, \alpha_2 \in \mathbb{C}$ are arbitrary.

We are going to obtain new solutions of the MTM system (\ref{MTM}) in the form (\ref{u-two-sol})--(\ref{v-two-sol}) and (\ref{MTM-double}) by using a new parameterization of the exponential two-soliton solutions (\ref{2-soliton}) and by taking the limits to the algebraic two-soliton solutions. 

\subsection{New parameterization of the exponential two-soliton solutions}

In order to represent the two-soliton solutions (\ref{2-soliton}) in the meaningful way where each soliton resembles the exponential soliton given by (\ref{MTM-solitary}), we will use the following parameterization: 
\begin{equation}
\label{par-2-solitons}
p_j = \mathrm{i} \delta_j e^{-i \gamma_j}, \quad 
\alpha_j = 2 \sqrt{\delta_j} \sin \gamma_j e^{\frac{\mathrm{i} \gamma_j}{2} + \sin \gamma_j x_j - i \cos \gamma_j t_j}, \quad j = 1,2,
\end{equation}
with arbitrary parameters $\gamma_j \in (0, \pi)$, $\delta_j > 0$, and $(x_j,t_j) \in \mathbb{R}^2$. By using the parameterization (\ref{par-2-solitons}) for $p_j$, we obtain 
$$
\zeta_j = \sin \gamma_j \left[ \frac{1}{2} (\delta_j + \delta_j^{-1}) x + \frac{1}{2} (\delta_j - \delta_j^{-1}) t \right] + \mathrm{i} \cos \gamma_j \left[ \frac{1}{2} (\delta_j - \delta_j^{-1}) x + \frac{1}{2} (\delta_j + \delta_j^{-1}) t \right].
$$
This representation resembles the Lorentz transformation (\ref{MTM-Lorentz}) with 
$$
\frac{1}{2} (\delta_j + \delta_j^{-1}) = \frac{1}{\sqrt{1-c_j^2}}, \quad 
\frac{1}{2} (\delta_j - \delta_j^{-1}) = \frac{c_j}{\sqrt{1-c_j^2}},
$$
where we have introduced the wave speeds
\begin{equation}
\label{wave-speed}
c_j := \frac{\delta_j^2-1}{\delta_j^2+1} \in (-1,1), \quad j = 1,2.
\end{equation}
Due to parameterization (\ref{par-2-solitons}), we 
obtain 
$$
c_{j j} =e^{-\mathrm{i} \gamma_j + 2 \sin \gamma_j x_j}, \quad j = 1,2,
$$
and, more generally, 
$$
c_{i j} =-\frac{4 \sqrt{\delta_i \delta_j} \sin \gamma_i \sin \gamma_j \delta_i}{(\delta_i e^{-\frac{\mathrm{i}}{2}(\gamma_i + \gamma_j)} - \delta_j e^{\frac{\mathrm{i}}{2} (\gamma_i + \gamma_j)})^2} 
e^{-\frac{\mathrm{i}}{2}(\gamma_i + \gamma_j) + \sin \gamma_i x_i + \sin \gamma_j x_j + \mathrm{i} \cos \gamma_i t_i - \mathrm{i} \cos \gamma_j t_j}, 
$$
so that we can introduce the following two real-valued 
coordinates
\begin{equation}
\label{coordinates}
\xi_j := \sin \gamma_j \left( \frac{x + c_j t}{\sqrt{1-c_j^2}} + x_j\right), \quad 
\eta_j := \cos \gamma_j \left( \frac{t + c_j x}{\sqrt{1-c_j^2}} + t_j\right),
\end{equation}
where $(x_j,t_j) \in \mathbb{R}^2$ play the role of translational parameters 
in (\ref{MTM-symm}). 

To derive the explicit expressions for $c_{12j}$ and $c_{1212}$, we use (\ref{par-2-solitons}) and obtain 
\begin{align*}
c_{121} = (p_1-p_2) \bar{p}_1 \left[\frac{\bar{\alpha}_2 c_{1 1}}{p_1 (p_2 + \bar{p}_1)} - \frac{\bar{\alpha}_1 c_{2 1}}{p_2 (p_1+ \bar{p}_1)} \right] = \frac{\mathrm{i} \bar{p}_1 |\alpha_1|^2 \bar{\alpha}_2 (p_1-p_2)^2}{(p_1 + \bar{p}_1)^2 (\bar{p}_1 + p_2)^2}
\end{align*}
that 
\begin{align*}
c_{121} = \frac{(p_1-p_2)^2}{(\bar{p}_1+p_2)^2} \bar{\alpha}_2 e^{i \gamma_1 + 2 \sin \gamma_1 x_1},  \quad 
c_{122} = \frac{(p_1-p_2)^2}{(p_1+\bar{p}_2)^2} \bar{\alpha}_1 e^{i \gamma_2 + 2 \sin \gamma_2 x_2}.
\end{align*}
Similarly, we obtain
\begin{align*}
c_{1212} = e^{-\mathrm{i} \gamma_1 -\mathrm{i} \gamma_2 + 2 \sin \gamma_1 x_1 + 2 \sin \gamma_2 x_2} A_{12},
\end{align*} 
where
\begin{align*}
A_{12} &= \frac{|p_1-p_2|^2}{(p_1 + \bar{p}_2) (p_2 + \bar{p}_1)} \left[ 1 - \frac{16 \delta_1^2 \delta_2^2 \sin^2 \gamma_1 \sin^2 \gamma_2}{(p_1 + \bar{p}_1) (p_2 + \bar{p}_2) (p_1 + \bar{p}_2) (\bar{p}_1 + p_2)} \right] \\
&= -\frac{|p_1-p_2|^2}{(p_1 + \bar{p}_2)^2 (p_2 + \bar{p}_1)^2} \left[ (\delta_1 e^{-\mathrm{i}\gamma_1} - \delta_2 e^{\mathrm{i}\gamma_2})(\delta_2 e^{-\mathrm{i}\gamma_2} - \delta_1 e^{\mathrm{i}\gamma_1}) + 4 \delta_1 \delta_2 \sin \gamma_1 \sin \gamma_2 \right] \\
&= \left( \frac{\delta_1^2 + \delta_2^2 - 2 \delta_1 \delta_2 \cos(\gamma_1 - \gamma_2)}{\delta_1^2 + \delta_2^2 - 2 \delta_1 \delta_2 \cos(\gamma_1 + \gamma_2)} \right)^2.
\end{align*} 
This representation allows us to rewrite the component $f$ of the two-soliton solution (\ref{2-soliton}) in the explicit form:
\begin{align}
f &= 1 + e^{2 \xi_1 - \mathrm{i} \gamma_1} + e^{2 \xi_2 - \mathrm{i} \gamma_2} 
+ A_{12} e^{2 \xi_1 + 2 \xi_2 - \mathrm{i} \gamma_1 -\mathrm{i} \gamma_2} - 4 \sqrt{\delta_1 \delta_2} \sin \gamma_1 \sin \gamma_2 
e^{\xi_1 + \xi_2 - \frac{\mathrm{i}}{2} \gamma_1 - \frac{\mathrm{i}}{2} \gamma_2} \notag \\
& \quad 
\times 
\left[ \frac{\delta_1 e^{\mathrm{i} (\eta_1 - \eta_2)}}{(\delta_1 e^{-\frac{\mathrm{i}}{2}(\gamma_1 + \gamma_2)}  - \delta_2  e^{\frac{\mathrm{i}}{2}(\gamma_1 + \gamma_2)})^2} +
\frac{\delta_2 e^{-\mathrm{i} (\eta_1 - \eta_2)}}{(\delta_1 e^{\frac{\mathrm{i}}{2}(\gamma_1 + \gamma_2)}  - \delta_2  e^{-\frac{\mathrm{i}}{2}(\gamma_1 + \gamma_2)})^2} \right], 
\label{two-sol-f}
\end{align}
where $\xi_j$ and $\eta_j$ are given by (\ref{coordinates}). The components $h$ and $g$ are written in the hybrid form for now:
\begin{align}
h &= \bar{\alpha}_1 e^{\zeta_1} \left[ 1 + \left( \frac{p_1 - p_2}{p_1 + \bar{p}_2} \right)^2 e^{2 \xi_2 + \mathrm{i} \gamma_2} \right] 
+ \bar{\alpha}_2 e^{\zeta_2} \left[ 1 + \left( \frac{p_1 - p_2}{\bar{p}_1 + p_2} \right)^2 e^{2 \xi_1 + \mathrm{i} \gamma_1} \right]
\label{two-sol-h}
\end{align}
and
\begin{align}
g &= \frac{\mathrm{i} \bar{\alpha}_1}{p_1} e^{\zeta_1} 
\left[ 1 + \left( \frac{p_1 - p_2}{p_1 + \bar{p}_2} \right)^2 e^{2 \xi_2 + 3 \mathrm{i} \gamma_2} \right] 
+\frac{\mathrm{i} \bar{\alpha}_2}{p_2} e^{\zeta_2} \left[ 1 + \left( \frac{p_1 - p_2}{\bar{p}_1 + p_2} \right)^2 e^{2 \xi_1 + 3 \mathrm{i} \gamma_1} \right].
\label{two-sol-g}
\end{align}
The two-soliton solution corresponds to two exponential solitons propagating according to their wave speeds $c_{1,2}$ obtained from $\delta_{1,2}$ by (\ref{wave-speed}) and having frequencies $\omega_{1,2} = \cos(\gamma_{1,2})$ obtained from $\gamma_{1,2}$. The one-soliton solution appears from this formula by taking $\xi_2 \to -\infty$:
\begin{align*}
u = \lim_{\xi_2 \to -\infty} \frac{g}{\bar{f}} 
= \frac{\mathrm{i} \bar{\alpha}_1 e^{\zeta_1} }{p_1 (1 + e^{2 \xi_1 + \mathrm{i} \gamma_1})} 
=  \sin \gamma_1 \delta_1^{-1/2} {\rm sech}\left( \xi_1 + \frac{\mathrm{i}}{2} \gamma_1 \right) e^{\mathrm{i} \eta_1}  
\end{align*}
and similarly, 
\begin{align*}
v = \lim_{\xi_2 \to -\infty}  \frac{h}{f} 
= \frac{\bar{\alpha}_1 e^{\zeta_1}}{1 + e^{2 \xi_1 - \mathrm{i} \gamma_1}} 
= \sin \gamma_1 \delta_1^{1/2} {\rm sech}\left( \xi_1 - \frac{\mathrm{i}}{2} \gamma_1 \right) e^{\mathrm{i} \eta_1},  
\end{align*}
from which we recognize the exact solution (\ref{MTM-solitary}) extended by  the symmetry transformations (\ref{MTM-symm}) and (\ref{MTM-Lorentz}).

\subsection{Limit to the algebraic two-soliton solutions}

Each soliton in the two-soliton solution expressed in the Hirota form (\ref{bilinear}) with (\ref{two-sol-f}), (\ref{two-sol-h}), (\ref{two-sol-g}) has four arbitrary parameters 
$\delta_j > 0$, $\gamma_j \in (0,\pi)$, and $(x_j,t_j) \in \mathbb{R}^2$ for $j = 1,2$. In order to get the algebraic two-soliton solutions, we need to take the limit $\gamma_j \to \pi$ for each $j = 1,2$. Hence, we set 
$$
\gamma_j=\pi-\epsilon_j, \quad j = 1,2
$$
and expand to the leading order
$$
\sin \gamma_j = \epsilon_j + \mathcal{O}(\epsilon_j^3), \quad \cos \gamma_j = 1 + \mathcal{O}(\epsilon_j^2). 
$$
We can then define 
$$
X_j := \frac{x + c_j t}{\sqrt{1-c_j^2}} + x_j, \quad 
T_j := \frac{t + c_j x}{\sqrt{1-c_j^2}} + t_j
$$
and expand 
\begin{align*}
\left( \frac{p_1-p_2}{p_1 + \bar{p}_2} \right)^2 = \left( \frac{\delta_1 e^{\mathrm{i} \epsilon_1} - \delta_2 e^{\mathrm{i} \epsilon_2}}{\delta_1 e^{\mathrm{i} \epsilon_1} - \delta_2 e^{-\mathrm{i} \epsilon_2}} \right)^2 = 1 - \frac{4 \mathrm{i} \epsilon_2 \delta_2}{\delta_1 - \delta_2} + \mathcal{O}(\epsilon_1^2,\epsilon_2^2)
\end{align*}
and
\begin{align*}
A_{12} = \left( \frac{\delta_1^2 + \delta_2^2 - 2 \delta_1 \delta_2 \cos(\epsilon_1 - \epsilon_2)}{\delta_1^2 + \delta_2^2 - 2 \delta_1 \delta_2 \cos(\epsilon_1 + \epsilon_2)} \right)^2 = 1 - \frac{8 \delta_1 \delta_2 \epsilon_1 \epsilon_2}{(\delta_1 - \delta_2)^2} + \mathcal{O}(\epsilon_1^2 \epsilon_2^2).
\end{align*} 
This yields the expansions:
\begin{align*}
f &= 1 - e^{\epsilon_1 (2 X_1 + \mathrm{i}) + \mathcal{O}(\epsilon_1^3)} - 
e^{\epsilon_2 (2 X_2 + \mathrm{i}) + \mathcal{O}(\epsilon_2^3)} 
+ A_{12} e^{\epsilon_1 (2 X_1 + \mathrm{i}) + \mathcal{O}(\epsilon_1^3) + 
\epsilon_2 (2 X_2 + \mathrm{i}) + \mathcal{O}(\epsilon_2^3)}  \\
& \quad 
+ 4 \sqrt{\delta_1 \delta_2} \epsilon_1 \epsilon_2 
\frac{\delta_1 e^{-\mathrm{i} (T_1 - T_2)} + \delta_2 e^{\mathrm{i} (T_1 - T_2)}}{(\delta_1 - \delta_2)^2} + \mathcal{O}(\epsilon_1^2 \epsilon_2,\epsilon_1 \epsilon_2^2), \\
h &= -2 \mathrm{i} \delta_1^{1/2} \epsilon_1 e^{-\mathrm{i} T_1} \left[ 1 - \left( \frac{p_1 - p_2}{p_1 + \bar{p}_2} \right)^2 e^{\epsilon_2 (2 X_2 - \mathrm{i}) + \mathcal{O}(\epsilon_2^3)} \right] 
\left[1 + \mathcal{O}(\epsilon_1^2) \right] \\
& \quad 
-2 \mathrm{i} \delta_2^{1/2} \epsilon_2 e^{-\mathrm{i} T_2} \left[ 1 + \left( \frac{p_1 - p_2}{\bar{p}_1 + p_2} \right)^2 e^{\epsilon_1 (2 X_1 - \mathrm{i}) + \mathcal{O}(\epsilon_1^3)}\right]
\left[1 + \mathcal{O}(\epsilon_2^2) \right], \\
g &= 2 \mathrm{i} \delta_1^{-1/2}  \epsilon_1 e^{-\mathrm{i} T_1} 
\left[ 1 - \left( \frac{p_1 - p_2}{p_1 + \bar{p}_2} \right)^2 e^{\epsilon_2 (2 X_2 - 3 \mathrm{i}) + \mathcal{O}(\epsilon_2^3)} \right] 
\left[1 + \mathcal{O}(\epsilon_1^2) \right] \\
& \quad + 2 \mathrm{i} \delta_2^{-1/2}  \epsilon_2 e^{-\mathrm{i} T_2} 
\left[ 1 - \left( \frac{p_1 - p_2}{\bar{p}_1 + p_2} \right)^2 e^{\epsilon_1 (2 X_1 - 3 \mathrm{i}) + \mathcal{O}(\epsilon_1^3)} \right] 
\left[1 + \mathcal{O}(\epsilon_2^2) \right].
\end{align*}
Hence, we get the power expansions
$$
f = \epsilon_1 \epsilon_2 F + \mathcal{O}(\epsilon_1^2\epsilon_2,\epsilon_1 \epsilon_2^2), \quad 
h = \epsilon_1 \epsilon_2 H + \mathcal{O}(\epsilon_1^2\epsilon_2,\epsilon_1 \epsilon_2^2), \quad 
g = \epsilon_1 \epsilon_2 G + \mathcal{O}(\epsilon_1^2\epsilon_2,\epsilon_1 \epsilon_2^2)
$$ 
with 
\begin{align}
\label{formula-F}
F = (2 X_1 + \mathrm{i}) (2 X_2 + \mathrm{i}) + \frac{4 \sqrt{\delta_1 \delta_2}}{(\delta_1 - \delta_2)^2} 
\left[ \sqrt{\delta_1} e^{-\frac{\mathrm{i}}{2} (T_1 - T_2)} - \sqrt{\delta_2} e^{\frac{\mathrm{i}}{2} (T_1 - T_2)} \right]^2,
\end{align}
\begin{align}
\label{formula-H}
H = 2 \mathrm{i} \delta_1^{1/2} e^{-\mathrm{i} T_1} \left[  2 X_2 - \mathrm{i} - \frac{4 \mathrm{i} \delta_2}{\delta_1 - \delta_2} \right] 
+ 2 \mathrm{i} \delta_2^{1/2} e^{-\mathrm{i} T_2} \left[  2 X_1 - \mathrm{i} + \frac{4 \mathrm{i} \delta_1}{\delta_1 - \delta_2} \right], 
\end{align}
and
\begin{align}
\label{formula-G}
G = -2 \mathrm{i} \delta_1^{-1/2} e^{-\mathrm{i} T_1} \left[  2 X_2 + \mathrm{i} - \frac{4 \mathrm{i} \delta_1}{\delta_1 - \delta_2} \right] 
- 2 \mathrm{i} \delta_2^{-1/2} e^{-\mathrm{i} T_2} \left[  2 X_1 + \mathrm{i} + \frac{4 \mathrm{i} \delta_2}{\delta_1 - \delta_2} \right].
\end{align}
The algebraic two-soliton solution of the MTM system (\ref{MTM}) appears in the Hirota form as 
\begin{equation}
\label{two-sol}
u = \frac{G}{\bar{F}}, \quad v = \frac{H}{F}
\end{equation}
and yields the exact solution (\ref{u-two-sol})--(\ref{v-two-sol}).
It describes two algebraic solitons traveling with the speeds $c_{1,2}$ obtained from $\delta_{1,2}$ by (\ref{wave-speed}).  A single algebraic solution appears by taking $X_2 \to \infty$:
\begin{align*}
u = \lim_{X_2 \to \infty} \frac{G}{\bar{F}} 
=  \frac{2 \delta_1^{-1/2}}{1 + 2 \mathrm{i} X_1} e^{-\mathrm{i} T_1}  
\end{align*}
and similarly, 
\begin{align*}
v = \lim_{X_2 \to \infty} \frac{H}{F} 
= \frac{2 \delta_1^{1/2}}{1 - 2 \mathrm{i} X_1} e^{-\mathrm{i} T_1}, 
\end{align*}
from which we recognize the exact solution (\ref{MTM-algebraic}) extended by 
the symmetry transformations (\ref{MTM-symm}) and (\ref{MTM-Lorentz}).

\subsection{Limit to the algebraic double-soliton}

Each algebraic soliton in the two-soliton solution expressed in the Hirota form (\ref{two-sol}) with (\ref{formula-F}), (\ref{formula-H}), and (\ref{formula-G}) has three arbitrary parameters 
$\delta_j > 0$ and $(x_j,t_j) \in \mathbb{R}^2$ for $j = 1,2$. We now take the limit $\delta_1 \to \delta_2$. Due to the Lorentz transformation (\ref{MTM-Lorentz}), it is sufficient to set 
$$
\delta_1 = 1 + \varepsilon, \quad \delta_2 = 1 - \varepsilon
$$
and take the limit $\varepsilon \to 0$. This choice gives the algebraic double-soliton with zero wave speed as in (\ref{MTM-double}). Expanding $X_{1,2}$ and $T_{1,2}$ in the first powers of $\varepsilon$, we write 
\begin{align*}
\left\{ \begin{array}{l} 
X_1 = x + \varepsilon t + \frac{1}{2} \varepsilon^2 (x-t) - \frac{1}{2} \varepsilon^3 (x-t) + x_1 + \mathcal{O}(\varepsilon^4), \\
X_2 = x - \varepsilon t + \frac{1}{2} \varepsilon^2 (x-t) + \frac{1}{2} \varepsilon^3 (x-t)  + x_2 + \mathcal{O}(\varepsilon^4), \\
T_1 = t + \varepsilon x - \frac{1}{2} \varepsilon^2 (x-t) + \frac{1}{2} \varepsilon^3 (x-t)  + t_1 + \mathcal{O}(\varepsilon^4), \\
T_2 = t - \varepsilon x - \frac{1}{2} \varepsilon^2 (x-t) - \frac{1}{2} \varepsilon^3 (x-t) + t_2 + \mathcal{O}(\varepsilon^4). \end{array} \right.
\end{align*}
In view of the translational symmetry (\ref{MTM-symm}), it is also sufficient 
to set 
\begin{align*}
\left\{ \begin{array}{l} 
x_1 = \varepsilon a_1 + \frac{1}{2} \varepsilon^2 a_2 - \frac{1}{2} \varepsilon^3 a_3, + \mathcal{O}(\varepsilon^4), \\
x_2 = -\varepsilon a_1 + \frac{1}{2} \varepsilon^2 a_2 + \frac{1}{2} \varepsilon^3 a_3, + \mathcal{O}(\varepsilon^4), \\
t_1 = \varepsilon b_1 - \frac{1}{2} \varepsilon^2 b_2 + \frac{1}{2} \varepsilon^3 b_3, + \mathcal{O}(\varepsilon^4), \\
t_2 = -\varepsilon b_1 - \frac{1}{2} \varepsilon^2 b_2 - \frac{1}{2} \varepsilon^3 b_3, + \mathcal{O}(\varepsilon^4),
\end{array} \right.
\end{align*}
with arbitrary parameters $a_1$, $a_2$, $a_3$, $b_1$, $b_2$, and $b_3$. This gives the algebraic double-soliton with zero translational parameters for $(x,t)$ as in (\ref{MTM-double}).

\underline{For expansion of $F$,} we use
\begin{align*}
(2X_1 + \mathrm{i})(2X_2 + \mathrm{i}) &= (2x + \mathrm{i})^2 + \varepsilon^2 [2 (2x+\mathrm{i})(x-t+a_2) - 4 (t+a_1)^2] + \mathcal{O}(\varepsilon^4)
\end{align*}
and
\begin{align*}
& \quad \sqrt{\delta_1} e^{-\frac{\mathrm{i}}{2} (T_1 - T_2)} - \sqrt{\delta_2} e^{\frac{\mathrm{i}}{2} (T_1 - T_2)} \\
&= -2 \mathrm{i} \left( 1 - \frac{\varepsilon^2}{8} \right) \sin\left( \varepsilon (x+b_1) + \frac{\varepsilon^3}{2} (x-t+b_3) \right) + 2 \left( \frac{\varepsilon}{2} + \frac{\varepsilon^3}{16} \right) \cos\left( \varepsilon (x+b_1) \right) + \mathcal{O}(\varepsilon^5) \\
&= \varepsilon (1 - 2 \mathrm{i} (x+b_1)) + \varepsilon^3 \left[ 
\frac{\mathrm{i}}{3} (x+b_1)^3 - \mathrm{i} (x - t + b_3) + \frac{\mathrm{i}}{4} (x+b_1) - \frac{1}{2} (x+b_1)^2 + \frac{1}{8} \right] + \mathcal{O}(\varepsilon^5).
\end{align*}
Substituting expansions into (\ref{formula-F}) yields 
$F = F_0 + \varepsilon^2 F_2 + \mathcal{O}(\varepsilon^4)$ with 
\begin{align*}
F_0 &= -4b_1 (2x + \mathrm{i} + b_1), \\
F_2 &= 2 (2x+\mathrm{i}) (x-t+a_2) - 4 (t + a_1)^2 - \frac{1}{2} (1 - 2 \mathrm{i} (x+b_1))^2 \\
& \qquad + 2 (1 - 2 \mathrm{i} (x+b_1)) \left[ 
\frac{\mathrm{i}}{3} (x+b_1)^3 - \mathrm{i} (x - t + b_3) + \frac{\mathrm{i}}{4} (x+b_1) - \frac{1}{2} (x+b_1)^2 + \frac{1}{8} \right].
\end{align*}
If $b_1 \neq 0$, then the limit $\varepsilon \to 0$ recovers a single algebraic soliton in the form (\ref{MTM-algebraic}). However, if $b_1 = 0$,
then we get 
\begin{align*}
F_2 & = (1-2  \mathrm{i} x) \left[ \frac{2  \mathrm{i}}{3} x^3 + 3  \mathrm{i} x - x^2 - \frac{1}{4} + 2  \mathrm{i} (a_2 - b_3) \right] - 4 (t + a_1)^2  \\
&= -\frac{1}{12} \left[ 3 - 24  \mathrm{i} x - 24  x^2 - 32  \mathrm{i} x^3 - 16 x^4 + 48 (t + a_1)^2 + 24 (b_3 - a_2) (2x +  \mathrm{i}) \right].
\end{align*}

\underline{For expansion of $H$,} we use 
\begin{align*}
& \quad \delta_1^{1/2} e^{- \mathrm{i} T_1} (2 X_2 -  \mathrm{i}) + \delta_2^{1/2} e^{- \mathrm{i} T_2} (2 X_1 -  \mathrm{i}) \\
&= e^{- \mathrm{i} t + \frac{\mathrm{i}}{2} \varepsilon^2 (x-t+b_2)} \left\{ 2(2x-\mathrm{i}) + \varepsilon^2 \left[ 2 (x-t+a_2) + 2 (t+a_1) (2  \mathrm{i} (x+b_1) -1) \right] \right. \\
& \qquad \left. - \varepsilon^2 (2x - \mathrm{i}) \left[ (x+b_1)^2 + \mathrm{i}(x+b_1) + \frac{1}{4} \right] \right\} + \mathcal{O}(\varepsilon^4)
\end{align*}
and
\begin{align*}
& \quad \delta_2^{1/2} e^{- \mathrm{i} T_1} - \delta_1^{1/2} e^{- \mathrm{i} T_2} \\
& = e^{- \mathrm{i} t + \frac{\mathrm{i}}{2} \varepsilon^2 (x-t+b_2)} \;
\left[ -2 \mathrm{i} \left( 1 - \frac{\varepsilon^2}{8} \right) \sin\left( \varepsilon (x+b_1) + \frac{\varepsilon^3}{2} (x-t+b_3) \right) \right. \\
& \qquad \qquad \left.
- 2 \left( \frac{\varepsilon}{2} + \frac{\varepsilon^3}{16} \right) \cos\left( \varepsilon (x+b_1) \right) + \mathcal{O}(\varepsilon^5)\right] \\
&= e^{- \mathrm{i} t + \frac{\mathrm{i}}{2} \varepsilon^2 (x-t+b_2)} 
\left\{ - \varepsilon (1 + 2 \mathrm{i} (x+b_1)) \right. \\
& \quad \left. 
+ \varepsilon^3 \left[ \frac{\mathrm{i}}{3} (x+b_1)^3 + \frac{\mathrm{i}}{4} (x+b_1) - \mathrm{i} (x-t+b_3) + \frac{1}{2} (x+b_1)^2 - \frac{1}{8} \right] 
+ \mathcal{O}(\varepsilon^5) \right\}.
\end{align*}
Substituting expansions into (\ref{formula-H}) yields 
$H = e^{- \mathrm{i} t + \frac{\mathrm{i}}{2} \varepsilon^2 (x-t+b_2)}  \left[ H_0 + \varepsilon^2 H_2 + \mathcal{O}(\varepsilon^4) \right]$ with 
\begin{align*}
H_0 &= -8 \mathrm{i} b_1, \\
H_2 &= 2 \mathrm{i} \left[2 (x-t+a_2) + 2 (t+a_1) (2  \mathrm{i} (x+b_1) -1) - 
 (2x - \mathrm{i}) \left[ (x+b_1)^2 + \mathrm{i}(x+b_1) + \frac{1}{4} \right] \right]\\
& \qquad + 4 \left[ \frac{\mathrm{i}}{3} (x+b_1)^3 + \frac{\mathrm{i}}{4} (x+b_1) - \mathrm{i} (x-t+b_3) + \frac{1}{2} (x+b_1)^2 - \frac{1}{8} \right] 
+ 2 \left[ 1 + 2 \mathrm{i}(x+b_1) \right].
\end{align*}
We confirm again that if $b_1 \neq 0$, then the limit $\varepsilon \to 0$ recovers a single algebraic soliton in the form (\ref{MTM-algebraic}). However, if $b_1 = 0$, then we get 
\begin{align*}
H_2 &= 2 \mathrm{i} \left[2 (a_2-b_3) + 2 (t+a_1) (2  \mathrm{i} x -1) - \frac{4}{3} x^3 - 2 \mathrm{i} x^2 + x - \frac{\mathrm{i}}{2} \right] \\
&= -\frac{1}{3} \left[ -3 - 6 \mathrm{i} x - 12 x^2 + 8 \mathrm{i} x^3 
+ 12 (t + a_1) (2x + \mathrm{i}) + 12 \mathrm{i} (b_3-a_2) \right]. 
\end{align*}

\underline{Similarly for $G$,} we obtain in the case of $b_1 = 0$ that $G = \varepsilon^2 e^{-\mathrm{i} t} G_2 + \mathcal{O}(\varepsilon^4)$ with 
\begin{align*}
G_2 = -\frac{1}{3} \left[ -3 + 6 \mathrm{i} x - 12 x^2 - 8 \mathrm{i} x^3 
- 12 (t + a_1) (2x - \mathrm{i}) - 12 \mathrm{i} (b_3-a_2) \right]. 
\end{align*}
The limit $\varepsilon \to 0$ in (\ref{two-sol}) yields a new solution for the algebraic double-soliton in the form: 
\begin{equation}
\label{MTM-degenerate}
\left[ \begin{matrix}
u(x,t) \\
v(x,t)  
\end{matrix} \right]  = 
\left[ \begin{matrix} \displaystyle  \frac{4(-3 + 6 \mathrm{i} x - 12 x^2 - 8 \mathrm{i} x^3 - 12 (t+ \alpha) (2x - \mathrm{i}) - \mathrm{i}\beta)}{3 + 24 \mathrm{i} x - 24 x^2 + 32 \mathrm{i} x^3 - 16 x^4 + 48 (t+\alpha)^2 + 2 \beta (2x - \mathrm{i}) } \\ \displaystyle  \frac{4(-3 - 6 \mathrm{i} x - 12 x^2 + 8 \mathrm{i} x^3 + 12 (t+ \alpha) (2x + \mathrm{i}) + \mathrm{i}\beta)}{3 - 24 \mathrm{i} x - 24 x^2 - 32 \mathrm{i} x^3 - 16 x^4 + 48 (t+\alpha)^2 + 2 \beta (2x + \mathrm{i}) } \end{matrix} \right]  e^{-\mathrm{i} t}.
\end{equation}
where $\alpha := a_1$ and $\beta := 12 (b_3 - a_2)$ are two real-valued parameters. Due to the symmetry transformation (\ref{MTM-symm}), the parameter $\alpha$ is trivial and can be set to $0$ as is done in (\ref{MTM-double}).

Note that we have confirmed validity of (\ref{MTM-degenerate}) by searching for polynomial solutions of the bilinear equations (\ref{Hirota}) with $f$ being polynomial in $x$ of degree $4$ and in $t$ of degree $2$  
and with $h$ and $g$ being polynomials in $x$ of degree $3$ and in $t$ of degree $1$. The only parameters of the polynomial solutions were found to be $\alpha, \beta \in \mathbb{R}$ as in (\ref{MTM-degenerate}).

\subsection{Mass of the algebraic double-soliton}

We shall prove (\ref{double-mass}) here. It follows from (\ref{bilinear}) and (\ref{Hirota}) that 
$$
|u|^2 + |v|^2 = \frac{|g|^2 + |h|^2}{|f|^2} = 2 \mathrm{i} \left( \frac{f_x}{f} - 
\frac{\bar{f}_x}{\bar{f}} \right),
$$
where 
$$
f = 16 x^4 + 32 \mathrm{i} x^3 + 24 x^2 + 24 \mathrm{i} x - 3 - 48 t^2 - 2 \beta (2x + \mathrm{i}).
$$
We claim that $f$ has no zeros on $\mathbb{R}$ in $x$ for every $t \in \mathbb{R}$ and $\beta \in \mathbb{R}$. Indeed, if $x,t,\beta \in \mathbb{R}$, zeros of $f$ must satisfy
\begin{equation*}
\left\{ \begin{array}{l} 
16 x^4 + 2 x^2 - 3 - 48 t^2 - 4 \beta x = 0, \\
32 x^3 + 24 x - 2 \beta = 0. \end{array}
\right.
\end{equation*}
Expressing $\beta = 16 x^3 + 12 x$ yields $-48 x^4 - 24 x^2 - 3 - 48 t^2 = 0$, 
which cannot be satisfied for $x,t \in \mathbb{R}$. Hence, there exist no roots of $f$ on $\mathbb{R}$ in $x$ for every $t \in \mathbb{R}$ and $\beta \in \mathbb{R}$. This and the fast decay at infinity,  
$$
\frac{f_x}{f} - 
\frac{\bar{f}_x}{\bar{f}} = \mathcal{O}\left(\frac{1}{|x|^2}\right) \quad \mbox{\rm as} \;\; |x| \to \infty,
$$
justify the applications of Jordan's lemma and the argument principle to 
compute the integral on $\mathbb{R}$ with techniques of complex analysis:
\begin{align*}
\int_{\mathbb{R}} (|u|^2 + |v|^2) dx &= \lim_{R \to \infty} 
\int_{[-R,R] \cup C_R^+} (|u|^2 + |v|^2) dz \\
&= 2 \mathrm{i} \lim_{R \to \infty} 
\int_{[-R,R] \cup C_R^+} \left( \frac{f_x}{f} - 
\frac{\bar{f}_x}{\bar{f}} \right) dz \\
&= 4 \pi (N_{\bar{f}} - N_f),
\end{align*}
where $C_R^+$ is a semicircle of radius $R$ in the upper half of the complex extension of $x$ denoted by $\mathbb{C}_+$,  $N_f$ is the number of zeros of $f$ in $\mathbb{C}_+$ and 
$N_{\bar{f}}$ is the number of zeros of $\bar{f}$ in $\mathbb{C}_+$. 
Since $f$ has no zeros on $\mathbb{R}$, we have 
$$
N_{\bar{f}} = \deg(f) - N_f. 
$$
Since $\deg(f) = 4$, we only need to show that $N_f = 1$ to obtain (\ref{double-mass}). However, this is true for every $\beta \in \mathbb{R}$ 
as $|t| \to \infty$ due to the representation of $f$ in the equivalent form 
$$
f = (2x+\mathrm{i})^4 + 12 (2x+\mathrm{i})^2 - 4 \mathrm{i} (2x+\mathrm{i}) - 2 \beta (2x+\mathrm{i}) + 4 - 48 t^2,
$$
from which we have 
$$
(2x+\mathrm{i}) = \sqrt[4]{12} \sqrt{|t|} e^{\frac{\mathrm{i} \pi n}{2}} + \mathcal{O}\left(\frac{1}{\sqrt{|t|}}\right) \quad \mbox{\rm as} \;\; |t| \to \infty,
$$
where $n = 0,1,2,3$. There is only one root in $\mathbb{C}_+$ which corresponds to $n = 1$. Since the number $N_f$ cannot change in the continuation of $f$ in $t \in \mathbb{R}$ for every $\beta \in \mathbb{R}$, we have $N_f = 1$ for every $t \in \mathbb{R}$ and $\beta \in \mathbb{R}$. Hence $Q(u_{\rm double},v_{\rm double}) = 8 \pi$ and (\ref{double-mass}) holds for every $\beta \in \mathbb{R}$.

\section{Conclusion}
\label{sec4}

We have constructed exact solutions of the MTM system (\ref{MTM}) for fast and slow scatterings of two algebraic solitons. Each algebraic soliton 
appears as the zero-energy resonance of the nonlinear Dirac equations \cite{Guo23} and is supported by the embedded eigenvalue in the KN spectral problem \cite{KPR06}. The fast scattering of two algebraic solitons with different wave speeds $c_1 \neq c_2$ is described by the exact soluton (\ref{u-two-sol}) and (\ref{v-two-sol}). The slow scattering of two identical solitons with zero speed is described by the exact solution (\ref{MTM-double}). The exact solutions suggest that the algebraic 
solitons are stable coherent structures arising in a more complicated evolution of the MTM system (\ref{MTM}). 

These discoveries lead to a number of open questions which can be addressed in future research. First, the mathematical problem of proving orbital stability of algebraic solitons is still open with only partial progress obtained within the derivative NLS equation in \cite{Wu18}. Second, the algebraic double-soliton solution (\ref{MTM-double}) suggests existence of a hierarchy 
of higher-order rational solutions of the MTM system (\ref{MTM}) which has not been obtained in the previous works \cite{Feng2,He,Ye}. Third, a similar algebraic double-soliton and a similar hierarchy of higher-order rational solutions must exist in the other nonlinear equations associated with the KN spectral problem, among which the most significant model is the derivative NLS equation \cite{Ling,HeAlg3}. Finally, development of the IST methods and the generalized Darboux transformation methods for the algebraic solitons associated with the embedded eigenvalues of the KN spectral problem is still a challenging problem for future research. 

\vspace{0.2cm}

{\bf Acknowledgements.} The authors thank B. F. Feng, L. Ling, and J. He for providing useful references. 

\vspace{0.2cm}

{\bf Conflict of Interest:} The authors have no conflict of interest.

\vspace{0.2cm}

{\bf Data availability statement:} The paper does not contain new numerical data.


\begin{thebibliography}{1}
	
\bibitem{AC90} M. J. Ablowitz and P. A. Clarkson, {\em Solitons, nonlinear evolution equations and inverse scattering} (Cambridge University Press, Cambridge, 1991).

\bibitem{Borreli} W. Borrelli, R. Carlone, and L. Tentarelli, 
``On the nonlinear Dirac equation on noncompact metric graphs", 
J. Diff. Eqs. {\bf 278} (2021) 326--357.

\bibitem{BC19} N. Boussaid and A. Comech, 
{\em Nonlinear Dirac equations: Spectral stability of solitary waves}, Mathematical Surveys and Monographs {\bf 244} (American Mathematical Society, Providence, RI, 2019).

\bibitem{CPS16} A. Contreras, D.E. Pelinovsky, and Y. Shimabukuro, \textit{ $L^2$ orbital stability of Dirac solitons in the massive Thirring model}, Commun. PDEs {\bf 41} (2016) 227--255.

\bibitem{Feng1} J. Chen and B.-F. Feng, 
 \textit{ Tau-function formulation for bright, dark soliton and breather solutions to the massive Thirring model}, Stud. Appl. Math. {\bf 150} (2023) 35--68.
 
\bibitem{Feng2} J. Chen, B. Yang, and B.-F. Feng, \textit{ Rogue waves in the massive Thirring model}, Stud Appl Math.  {\bf 151} (2023) 1020--1052.
 
\bibitem{Akh16} A. Chowdury, A. Ankiewicz, and N. Akhmediev, \textit{ Periodic and rational solutions of modified Korteweg-de Vries equation}, 
Eur. Phys. J. D {\bf 70} (2016) 104.

\bibitem{Guo21} Y. Ding, X. Dong, and Q. Guo, \textit{ Nonrelativistic limit and some properties of solutions for nonlinear Dirac equations}
Calc. Var. PDEs {\bf 60} (2021) 144 (23 pp.)

\bibitem{GPS93} K.A. Gorshkov, D.E. Pelinovsky, and Yu.A. Stepanyants, \textit{ Normal and anomalous scattering, formation and decay of bound states of two--dimensional solitons described by the Kadomtsev--Petviashvili equation}, JETF {\bf 77} (1993) 237--245

\bibitem{Ling} B. Guo, L. Ling, and Q. P. Liu, \textit{ High--order solutions and generalized Darboux transformations of derivative nonlinear Schr\"{o}dinger
Equations}, Stud. Appl. Math. {\bf 130} (2013) 317--344.

\bibitem{He} L. Guo, L. Wang, Y. Cheng, and J. He, \textit{ High-order rogue wave solutions of the classical massive Thirring model equations}, 
Commun. Nonlinear Sci. Numer. Simulat. {\bf 52} (2017) 11--23.

\bibitem{Guo23} Q. Guo and Y. Ke, \textit{ Solitary solutions of a nonlinear Dirac equation with different frequencies}, arXiv:2310.19132 (2023)

\bibitem{Cheng23} C. He, J. Liu, and C. Qu, \textit{ Massive Thirring model: inverse scattering and soliton resolution}, arXiv: 2307.15323 (2023)

\bibitem{S18} W. Hu, W. Huang, Z. Lu, and Y. Stepanyants, \textit{ Interaction of multi-lumps within the Kadomtsev--Petviashvili equation}, Wave Motion {\bf 77} (2018) 243--256.

\bibitem{KN76} D.J. Kaup and A.C. Newell, \textit{ On the Coleman correspondence and the solution of the massive Thirring model}, Lett. Nuovo Cimento {\bf 20} (1977) 325--331.
  
\bibitem{KPR06} M. Klaus, D.E. Pelinovsky, and V.M. Rothos,  \textit{ Evans function for Lax operators with algebraically decaying potentials}, J. Nonlin. Science {\bf 16} (2006) 1--44.

\bibitem{KM77} E.A. Kuznetsov and A.V. Mikhailov, \textit{On the complete integrability of the two-dimensional classical Thirring model}, Theor. Math. Phys. {\bf 30} (1977) 193--200.

\bibitem{Wu18} S. Kwon and Y. Wu, \textit{ Orbital stability of solitary waves for derivative nonlinear Schr\"{o}dinger equation}, 
J. Anal. Math. {\bf 135} (2018)  473--486.

\bibitem{LiPel} Z. Li and D. Pelinovsky, \textit{ Double-soliton solutions of the massive Thirring model}, preprint (2024)

 \bibitem{Pel11}
 D. E. Pelinovsky {\sl{ Localization in Periodic Potentials; from Schrödinger Operators to the Gross--Pitaevskii Equation}}, London Mathematical Society Lecture Note Series, 2011. 
 
 \bibitem{PG97} D.E. Pelinovsky and R.H.J. Grimshaw, \textit{ Structural transformation of eigenvalues for a perturbed algebraic soliton potential}, Phys. Lett. A {\bf 229} (1997) 165--172.
 
 \bibitem{PS19} D.E. Pelinovsky and A. Saalmann, \textit{ Inverse scattering for the massive Thirring model} in {\em Nonlinear Dispersive Partial Differential Equations and Inverse Scattering} (Editors: P. Miller, P. Perry, J.C. Saut, and C. Sulem), Fields Institute Communications {\bf 83} (Springer, New York, NY, 2019) 497--528.
 
 \bibitem{PS14} D.E. Pelinovsky and Y. Shimabukuro, \textit{ Orbital stability of Dirac solitons}, Lett. Math. Phys. {\bf 104} (2014) 21--41.
 
 \bibitem{HeAlg1}  Q. Xing, L. Wang, D. Mihalache, K. Porsezian, and J. He,
\textit{ Construction of rational solutions of the real modified Korteweg-de Vries
 equation from its periodic solutions}, Chaos {\bf 27} (2017) 053102 (14 pages).
 
 \bibitem{HeAlg2}  S. Xu, J. He, and L. Wang
\textit{ The Darboux transformation of the derivative
 nonlinear Schr\"{o}dinger equation}, 
 J. Phys. A: Math. Theor. {\bf 44} (2011) 305203 (22 pages).

 
\bibitem{Yang1} B. Yang and J. Yang, \textit{ Rogue wave patterns in the nonlinear Schr\"{o}dinger equation}, Physica D {\bf 419} (2021) 132850.

\bibitem{Yang2} B. Yang and J. Yang, \textit{  Universal rogue wave patterns associated with the Yablonskii–Vorob’ev polynomial hierarchy}, 
Physica D {\bf 425} (2021) 132958.

\bibitem{Yang3} B. Yang and J. Yang, \textit{ Rogue wave patterns associated with Okamoto polynomial hierarchies}, Stud. Appl. Math. {\bf 151} (2023) 60--115.
 
\bibitem{Ye} Y. L. Ye, L. L. Bu, C. C. Pan, S. H. Chen, D. Mihalache, and F. Baronio, \textit{ Super rogue wave states in the classical massive
Thirring model system}, Rom Rep Phys. {\bf 73} (2021) 117.

\bibitem{Yan} G. Zhang and Z. Yan, \textit{ The derivative nonlinear Schr\"{o}dinger equation with zero/nonzero boundary conditions: inverse scattering transforms and $N$-double-pole solutions}, J. Nonlin. Sci. {\bf 30} (2020) 3089--3127.

\bibitem{HeAlg3} Y. Zhang, L. Guo, S. Xu, Z. Wu, and J. He, 
\textit{ The hierarchy of higher order solutions of the derivative
	nonlinear Schr\"{o}dinger equation}, Commun Nonlinear Sci Numer Simulat 
{\bf 19} (2014) 1706--1722.
	
\bibitem{He2} Y. Zhang, D. Qiu, S. Shen, and J. He, \textit{ The revised Riemann--Hilbert approach to the Kaup--Newell equation with a non-vanishing boundary condition: simple poles and higher-order poles}, preprint (2024).

\bibitem{He1} Y. Zhang, J. Rao, Y. Cheng, and J. He, \textit{ Riemann--Hilbert method for the Wadati--Konno--Ichikawa equation:  $N$  simple poles and one higher-order pole}, Physica D {\bf 399} (2019) 173--185. 

\bibitem{Zhang} Y. Zhang, H. Wu, and D. Qiu, \textit{ Revised Riemann--Hilbert problem for the derivative nonlinear Schrödinger equation: Vanishing boundary condition}, Theor. Math. Phys. {\bf 217} (2023) 1595--1608.





\end{thebibliography}
\end{document}